\documentstyle[preprint,floats,pra,aps]{revtex}

\def\be{\begin{equation}}
\def\ee{\end{equation}}
\def\bea{\begin{eqnarray}}
\def\eea{\end{eqnarray}}
\font\extra=msbm10 scaled \magstep1
\def\bbb #1{\hbox{{\extra #1}}}

\tightenlines 
\begin{document}

\title{\LARGE \bf Refined Factorizations \\
 of Solvable Potentials}

\author{J. Negro, L.M.~Nieto and  O. Rosas--Ortiz\footnote{On leave of
         absence from
         {\it Departamento de F\'\i sica\/}, CINVESTAV-IPN,
         {\it A.P. 14-740, 07000 M\'exico~D.F., Mexico\/}.}}
\address{Departamento de F\'\i sica Te\'orica, 
Universidad de Valladolid\\
47011 Valladolid, Spain}

\maketitle

\begin{abstract}
A generalization of the factorization technique is shown to be a powerful
algebraic tool to discover further properties of a class of integrable
systems in Quantum Mechanics.  The method is applied in the study of
radial oscillator, Morse and Coulomb potentials to obtain a wide set of
raising and lowering operators, and to show clearly the connection that link
these systems. 
\end{abstract}


\section{Introduction}

We shall begin this section by recalling some basic facts of the
standard factorization method, as can be found for instance in
\cite{Miller,Infeld}, mainly to fix the notation. Afterwards, we will set up
the general lines to define more general factorizations, and the way they
depart from the conventional ones previously characterized.

Let us consider a sequence of
stationary one dimensional Schr\"odinger equations, labeled by an integer 
number
$\ell$, written in the form 
\be
H^{\ell} \psi_n^{\ell} \equiv \left\{ -\frac{d^2\,}{dx^2} +
V^{\ell}(x)\right\} \psi_n^{\ell}(x) =
E_n^{\ell} \psi_n^{\ell}(x),
\ee
where the constants $\hbar$ and $m$ have been conveniently reabsorved. 
If such a set (or `hierarchy') of Hamiltonians can be expressed as
\be
\label{1.1}
H^{\ell} = X^+_{\ell} X^-_{\ell} - q(\ell) =
X^-_{\ell{-}1} X^+_{\ell{-}1} - q(\ell{-}1),
\ee
where
\be
\label{1.2}
X^\pm_{\ell} = \mp\frac{d\,}{dx} + w_{\ell}(x),
\ee
being $w_{\ell}(x)$ functions
and $q(\ell)$ constants, then we will say that they admit a factorization.
From (\ref{1.2}) we have that $X^\pm_{\ell}$ are hermitian conjugated of 
each other, $\left(X^-_{\ell}\right)^{\dag} =X^+_{\ell}$, with respect to the
usual inner product of the Schr\"odinger equation. This is consistent with
the factorization (\ref{1.1}) and the hermiticity of $H^{\ell}$.

We shall focus our interest in studying the discrete spectrum of each 
Hamiltonian, so we further impose that the equation 
\be
\label{1.2a}
X^-_{\ell} \, \psi_{\ell}^{\ell}(x) = 0
\ee
will determine the ground state of $H^{\ell}$ if it exists. Of
course many  other properties related with the continuous  spectrum can
also be derived with the help of factorizations, but they are out of our 
present scope.

Some consequences that can immediately be derived from the previous 
conditions are enumerated below.

\begin{itemize}
\item[{\bf i)}] 
{\bf  Spectrum}. Let $\psi_{\ell}^{\ell}$ be the ground state of $H^{\ell}$
as stated in (\ref{1.2a}), then its energy is precisely
$E_{\ell}^{\ell} =  -q(\ell)$. When there are excited bounded states
$\psi^{\ell}_{n}$, with $n=\ell, \ell+1,\dots$, their energy  is
given by $E^{\ell}_n = -q(n)$. Therefore, in these circunstances, $-q(\ell)$
should be an increasing function on $\ell$.  

\item[{\bf ii)}] 
{\bf  Eigenfunctions}. It is straightforward to check, for each $\ell$, the
intertwining relations
\be
\label{1.2b}
H^{\ell} X^+_{\ell} = X^+_{\ell} H^{\ell+1}, \qquad 
 H^{\ell+1} X^-_{\ell}=X^-_{\ell} H^{\ell}. 
\ee 
Let us designate by  
${\cal H}^{\ell}= \langle \{\psi_{n}^{\ell}\}_{n\geq\ell}\rangle$ the Hilbert
space spanned by the bounded states of $H^{\ell}$, for $\ell\in {\bbb Z}$.
Then, due to (\ref{1.2b}) the operators $X^\pm_{\ell}$ link these spaces as 
\be \label{accion}
\begin{array}{cc}
X^-_{\ell}:{\cal H}^{\ell} \to {\cal H}^{\ell+1} \qquad
&X^+_{\ell}:{\cal H}^{\ell+1} \to {\cal H}^{\ell} \\ [1ex]
X^-_{\ell}\psi_{n}^{\ell}(x) \propto \psi_{n}^{\ell+1}(x),\qquad
&X^+_{\ell}\psi_{n}^{\ell+1}(x) \propto \psi_{n}^{\ell}(x).
\end{array}
\ee
\end{itemize}
Remark that the action of $X^\pm_{\ell}$ preserve the label $n$, that is,
they connect eigenfunctions with the same energy $E^{\ell}_n$. If the
eigenfunctions are normalized we can be more explicit: up to an arbitrary
phase factor,
\be
\label{accB}
\begin{array}{ll}
X^-_{\ell}\, \psi_{n}^{\ell}(x)=
{\sqrt{q(\ell){-}q(n)}}\,
\psi_{n}^{\ell+1}(x),  & n \geq \ell \\ [1ex]
X^+_{\ell}\, \psi_{n}^{\ell+1}(x)= 
{\sqrt{q(\ell){-}q(n)}}\,
\psi_{n}^{\ell}(x),\qquad  &n > \ell.
\end{array}
\ee

Similar considerations would also apply if the ground states were defined
through $X^+$. Depending on each particular problem we will use one of
the following notations
\bea
\label{xmas}
&&  X^+_{\ell-1}(r) \, \psi_{-\ell}^{\ell} = 0, \  \  {\rm if} \ \ \ell\leq 0,
\\ 
\label{xmasB}
&& X^+_{\ell-1}(r) \, \psi_{\ell}^{\ell} = 0, \  \  {\rm if} \ \ \ell\geq 0.
\eea
For such a case $-q(\ell)$ must be a decreasing function of $\ell$. We shall
also have the opportunity to illustrate this situation in some examples along 
the next sections.  

Now, it is natural to define a set of free-index linear operators 
$\{X^\pm,L\}$ acting on the direct sum of the Hilbert spaces
${\cal H}\equiv {\displaystyle \oplus_{\ell} {\cal H}^{\ell} }$ by means of
\be
\label{free}
X^-\psi_{n}^{\ell}:=X^-_{\ell}\psi_{n}^{\ell},\qquad
X^+\psi_{n}^{\ell}:=X^+_{\ell-1}\psi_{n}^{\ell},\qquad
L \psi_{n}^{\ell}:= \ell \,  \psi_{n}^{\ell},
\ee
where one must have in mind (\ref{accion}) and (\ref{accB}). That is, the
operators $X^{\pm}$ act on each function $\psi_{n}^{\ell}(x)$ by means of 
the differential operators (\ref{1.2}) changing $\ell$ into
$\ell\mp 1$.
The action on any
other vector of ${\cal H}$ can be obtained from (\ref{free}) by linearization,
but we shall never need it. At this moment we are not in conditions to 
guarantee that the space ${\cal H}$ is invariant under this action
(it might happen that the action of $X^\pm$ on ${\cal H}$ could lead us to 
the continuous spectrum, or even to an unphysical eigenfunction), but we
postpone this problem to the examples of Section III.

Taking into account our definitions (\ref{free}), it is straightforward to 
arrive at the following commutators,
\be \label{com}
[L,X^\pm]=\mp X^\pm,\qquad
[X^+,X^-] = q(L)-q(L{-}1).
\ee
It is clear that the set of operators $\{X^\pm,L\}$ in general does not close a
Lie algebra; relations (\ref{com}) only allow us  to speak formally of an
associative algebra. 

There are many aspects of the conventional factorizations above 
characterized which can be modified, mainly with the objective of being 
aplicable to a wider class of systems (see for example \cite{Mielnik}).
However, in this paper we are interested in going deeply into the
possibilities of this method on a class of systems where the usual
factorization can already be applied, so that it could supply us with
additional information. With this aim, we shall stress here on two points 
that will be useful in the next sections. 

First, we shall assume that the operators $X_{\ell}^\pm$ do
not have to take necessarily the form given in (\ref{1.2}). In particular, if
we have a family of invertible operators $D_{\ell}$ and define 
$Y^+_{\ell} = X^+_{\ell}D_{\ell}^{-1}$, $Y^-_{\ell} = D_{\ell}X^-_{\ell}$,
we will also have
\be
H^{\ell} = X^+_{\ell} X^-_{\ell} - q(\ell) =
Y^+_{\ell} Y^-_{\ell} - q(\ell).
\ee
The new factor $D_{\ell}$ may be a function (which would
add nothing specially new) but also a local operator, i.e., an operator
acting on wavefunctions in the form
\be
D_{\ell}\,  \psi(x) =  \psi(g_{\ell}(x)),
\ee
where $g_{\ell}$ is a bijective real function. An
example of such an operator, which was already used in \cite{spiri}, is
given  by the dilation,
\be
\label{dil}
D(\mu) \psi(x) = \psi(\mu x),\qquad \mu>0.
\ee

Second, an eigenvalue equation can be characterized by more than one  
label; this consideration has also been explored by Barut {\it et al}
\cite{Barut},  but in another context. In the next section we shall deal with
two real labels; this will enable us to have more possible ways to
factorize the Hamiltonian hierarchy, and the sequence of labels will not be 
limited (essentialy) to the integers, but it will be constituted by a lattice of
points in
${\bbb R}^2$. This increasing of factorizations will reflect itself in a larger
algebra of free-index operators. In particular, among them, there can be
lowering and raising operators for each Hamiltonian, which can never be
obtained by the conventional factorization method. Section III will
illustrate how our general method  works when it is applied to three well
known potentials: radial oscillator, Morse, and radial Coulomb. For each of
these potentials we shall see that the results so obtained can be used to
recover, as special cases, those corresponding to the standard factorizations.
Finally some comments and remarks will end this paper.


\section{Refined Factorizations}

Once the spectrum $E^{\ell}_n$ of the hierarchy $H^{\ell}$ is known, we
propose a somewhat more general factorization of the eigenvalue equations 
than that one already displayed in (\ref{1.1}), as follows:
\be
h_{{ n},{ \ell}}(x) \left[ H^{{ \ell}} -
E^{\ell}_{{ n}} \right] =
B_{n,\ell} A_{n,\ell}- \phi(n,\ell) =
A_{\tilde n,\tilde \ell} B_{\tilde n,\tilde \ell}  - 
\phi(\tilde n,\tilde \ell) .
 \label{2.1}
\ee
This must be understood as a series of relationships valid for a class of
allowed values of the parameters $(n,\ell)\in {\bbb R}^2$. Here
$B_{n,\ell}$ and $A_{n,\ell}$ are first order differential operators in the
wider sense specified in the previous section,
$ h_{n,\ell}(x)$ denote functions, and
$\phi(n,\ell)$ are constants. The $({\widetilde n},{\widetilde \ell})$ values
depend on $(n,\ell)$, i.e., $({\widetilde n},{\widetilde \ell}) =  F(n,\ell)$,
being $F:{\bbb R}^2 \to {\bbb R}^2$ an invertible map defined on a certain
domain.  The iterated action of $F$ or $F^{-1}$ on a
fixed initial point $(n_0,\ell_0)\in  {\bbb R}^2$  originates a sequence of
points in ${\bbb R}^2$ that will play a role similar to the integer sequence
$\ell$ in ${\bbb R}$ for the usual factorizations. In principle the points
$(n,\ell)$ obtained by this new approach can take integer
values for both arguments, but we do not discard {\it a priori\/} other
possibilities. 

The problem of finding solutions to this kind of
factorizations becomes more involved because we have additional functions
$h_{n,\ell}(x)$ to be determined. Nevertheless, an important and immediate
consequence of (\ref{2.1}) is that the operators $B_{n,\ell},A_{n,\ell}$ share
properties similar to (\ref{1.2b})
with respect to their analogs $\{X_{\ell}^\pm\}$  :
\be
\begin{array}{l}
\left[h_{{\hat n},{\hat \ell}}(x) \left( H^{{\hat \ell}} 
- E^{\hat \ell}_{{\hat n}}
\right)\right]\, A_{n,\ell} = 
A_{n,\ell}\,
\left[ h_{{ n},{ \ell}}(x) \left( H^{{ \ell}} - E^{\ell}_{{ n}}\right)\right],
\\ [1.5ex]
B_{n,\ell}\, \left[h_{{\hat n},{\hat \ell}}(x) \left( H^{{\hat \ell}} 
- E^{\hat \ell}_{{\hat n}}
\right)\right] = 
\left[ h_{{ n},{ \ell}}(x) \left( H^{{ \ell}} - E^{\ell}_{{ n}}\right)\right]\,
B_{n,\ell}\, ,
\end{array}
\ee
where $F({\hat n},{\hat \ell}) = (n,\ell)$. Therefore, using the same
notation as in (\ref{accion}), 
\be
\label{2.2}
\begin{array}{cc}
A_{n,\ell}:{\cal H}^{\ell} \to {\cal H}^{{\hat \ell}}\qquad
&B_{n,\ell}:{\cal H}^{{\hat \ell}} \to {\cal H}^{\ell}\\
A_{n,\ell}\, \psi_{n}^{\ell}(x) \propto 
\psi_{{\hat n}}^{{\hat \ell}}(x),\qquad
&B_{n,\ell}\, \psi_{{\hat n}}^{{\hat \ell}}(x) \propto
\psi_{n}^{\ell}(x).
\end{array}
\ee
In this case, the most relevant differences with respect to the usual
factorizations are: 

\begin{itemize}
\item[i)] 
$B_{n,\ell},A_{n,\ell}$ in general do not preserve
the energy eigenvalue, they may change both labels $n$ and $\ell$. 

\item[ii)] 
$A_{n,\ell}$ does not act on the whole space ${\cal H}^{\ell}$, it acts just
on the eigenfunction $\psi_{n}^{\ell}(x)\in {\cal H}^{\ell}$ (the same can be
said of
$B_{n,\ell}$ with respect to $\psi_{{\hat n}}^{{\hat \ell}}(x) \in 
{\cal H}^{{\hat \ell}}$). 
\end{itemize}

When  $n= {\hat n}$ and $h_{n,\ell}(x) =1$, we
recover the conventional case with $B_{n,\ell}$, $A_{n,\ell}$ playing the
role of $X_{\ell}^+$, $X_{\ell}^-$, respectively. However, the hermiticity
properties for the general case are lost because the product
$B_{n,\ell}A_{n,\ell}$ gives not the Hamiltonian operator alone, but it
includes also a non constant multiplicative factor. 

We can define the free-index operators $\{A,B,L,N\}$ as we did in
(\ref{free}), where the latter is defined by 
$N \psi_{n}^{\ell}= n\, \psi_{n}^{\ell}$. They satisfy the following
commutation rules  
\be  \label{gencommuting} 
\begin{array}{lll}
[L,B] = B({\widetilde L}-L), \quad  & [N,B] = B({\widetilde N}-N),\quad
&[B,A] = \phi(N,L) - \phi({\widetilde N},{\widetilde L})
\\ [1ex]
[L,A] =  (L-{\widetilde L})A, \quad & [N,A]
=(N-{\widetilde N})A,\quad  &   [N,L] = 0,
\end{array}
\ee
where $({\widetilde N},{\widetilde L}) =F(N,L)$. As the operators $L,N$ 
commute, their eigenvalues are used to label the common eigenfunctions
$\psi_{n}^{\ell}(x)$.  We must also notice that the equation 
$A_{n,\ell} \psi_{n}^{\ell}(x)=0$ 
(or $B_{n,\ell} \psi_{{\hat n}}^{{\hat \ell}}(x)=0$)
does not necessarily give an eigenfunction  of $H^{\ell}$ (or $H^{\hat \ell}$);
this happens to be the case only when $\phi(n,\ell)=0$.


\section{Applications}

\subsection{Radial Oscillator Potential}

As usual the Hamiltonian of the two dimensional 
harmonic oscillator includes the effective radial potential
$V^{\ell}(r)=   r^2 +  \frac{(2\ell +1)(2\ell -1)}{4 r^2}$, where
$\ell=0,1\dots$ is for the angular momentun.
The related  stationary Schr\"odinger equation has discrete
eigenvalues denoted according to the following convention,  
$$
E_n^{\ell} = 2n +2,\qquad 
n=2\nu + \ell; \qquad \nu =0,1,\dots
$$ 
It can be factorized in two ways according to our general
scheme:
\be
\frac{-1}4 \left[ H^{ \ell} -
E^{\ell}_{ n} \right] = \frac{1}4  
\left[ \frac{d^2}{dr^2}  {-} r^2 {-}  \frac{(2\ell +1)(2\ell -1)}{4 r^2}  +
E^{\ell}_n \right] =  { B}^i_{ n, \ell}{ A}^i_{n,\ell}
 - \phi^i(n,\ell), \quad  i=1,2
\ee 
with $\phi^i(n,\ell)$ given by
\be
\phi^1(n,\ell) = -\frac12(n+\ell+2),\qquad
\phi^2(n,\ell) = -\frac12(n-\ell+2),
\ee
and where the action on the parameters associated to each factorization is 
given respectively by the functions 
\be
(n,\ell)= F^1(n+1,\ell+1),\qquad
(n,\ell)= F^2(n+1,\ell-1).
\ee
This can also be written in an easier notation,
\be
\label{otra}
\begin{array}{l}
A^1:{(n,\ell)} \to {(n+1,\ell+1)}\\ [1ex]
B^1:{(n+1,\ell+1)} \to {(n,\ell)}
\end{array}
\qquad
\begin{array}{l}
A^2:{(n,\ell)} \to {(n+1,\ell-1)}\\ [1ex]
B^2:{(n+1,\ell-1)} \to {(n,\ell)} .
\end{array}
\ee
The explicit form of these intertwining operators is 
\be
\label{explicitY}
\left\{
\begin{array}{l}
{ A}^1_{n,\ell}(r) = 
\frac12\left[\frac{d\ }{dr} {-} r {-}(\ell{+}1/2)\frac1r\right] \\[2ex]
{ B}^1_{n,\ell}(r) = 
\frac12\left[\frac{d\ }{dr} {+} r {+}(\ell{+}1/2)\frac1r\right] 
\end{array}
\right.
\qquad 
\left\{
\begin{array}{l}
{ A}^2_{n,\ell}(r) = 
\frac12\left[\frac{d\ }{dr} {-} r {+} (\ell{-}1/2)\frac1r\right] \\[2ex]
{ B}^2_{n,\ell}(r) = 
\frac12\left[\frac{d\ }{dr} {+} r {-} (\ell{-}1/2)\frac1r\right] 
\end{array}
\right.
\ee
Observe that in this case, as $h_{n,\ell}(r)$ is a constant, we are able
to implement also the hermiticity properties $(A^i)^{\dagger} = -B^i$. The
nonvanishing commutation rules for the free-index operators $\{N,L,A^i,B^i;
i=1,2\}$ are shown to be, in agreement with (\ref{gencommuting}), 
\be
\label{2boson}
\begin{array}{lll}
[L,B^i] =  (-1)^i B^i, \quad  & [N,B^i] =-B^i,\quad
&[A^i,B^i] = 1,
\\ [1ex]
[L,A^i] = -(-1)^i A^i, \quad & [N,A^i] = A^i, \quad  & i=1,2.   
\end{array}
\ee
These commutators correspond to two independent boson algebras with 
$N,L$ being a linear combination of their number operators. Formally we can
extend the values of $\ell$ so to include the negative integers. This is
physically appealing because in two space dimensions (only!)  $\ell$
represents the $L_z$-component of angular momentum, so that it could take
negative integer values. Of course the extension
$\psi^{-\ell}_{n}(r) := \psi^{\ell}_{n}(r)$ above proposed is consistent with 
such an interpretation: (i)  The radial components for opposite $L_z$-values
have to coincide, and (ii) The potential $V^{\ell}$ is invariant under the
interchange
$\ell \to -\ell$. With this convention, the Hilbert space
${\cal H}$ of bounded states is invariant under the action of the operators
$\{N,L,A^i,B^i; i=1,2\}$, so that it constitutes the support for a lowest weight
irreducible representation for the algebra (\ref{2boson}) based on the 
fudamental state
$\psi^{\ell=0}_{n=0}$.

It is worth to notice that, taking into account (\ref{otra}), the composition
$\{A^1A^2,B^1B^2\}$ constitutes the lowering and raising operators for each
Hamiltonian $H^{\ell}$, while the pair $\{A^1B^2,A^2B^1\}$ connects states
of different Hamiltonians $H^{\ell}$  with the same energy, changing only
the label $\ell$.

We shall compare briefly the above results with the conventional 
factorizations of the two-dimensional radial oscillator potential
\cite{Fernandez}. It is well known that there are two such factorizations
which we will write in the form:
\bea
(a)\quad
&& X_{\ell}^+X_{\ell}^- - q_x(\ell) = { H}_x^{\ell} =
{ H}^{\ell} - 2\ell \\  [1ex]
(b)\quad
&& Z_{\ell}^+Z_{\ell}^- - q_z(\ell) = { H}_z^{\ell} =
{ H}^{\ell} + 2\ell, 
\eea
with ${ H}^{\ell}=
-\frac{d^2}{dr^2}  {+} r^2 {+}  \frac{(2\ell +1)(2\ell -1)}{4 r^2}$. Then we 
have the following identification: 
\medskip

\noindent
Case $(a)$
\begin{enumerate}
\item
Operators: $X_{\ell}^+ = -2 { B}^1_{n,\ell}$, $X_{\ell}^- = 2 { A}^1_{n,\ell}$, 
$q_x(\ell)= 4\ell -2$.
\item
Ground states: $X_{\ell-1}^+ \psi^{\ell}_{-\ell}=0$, $\ell \leq 0$.
\item
Energy eigenvalues: $E_n^{\ell} = 4n+2$, with $n\in {\bbb Z}^+$ and $n\geq
-\ell$.
\end{enumerate}
In this case we have used a notation in agreement with (\ref{xmas}).
\medskip

\noindent
Case $(b)$
\begin{enumerate}
\item
Operators: $Z_{\ell}^+ = -2 { A}^2_{n-1,\ell+1}$, 
$Z_{\ell}^- = 2 {B}^2_{n-1,\ell+1}$, 
$q_z(\ell)= -4\ell -2$.
\item
Ground states: $Z_{\ell}^- \psi^{\ell}_{\ell}=0$, $\ell \geq 0$.
\item
Energy eigenvalues: $E_n^{\ell} = 4n+2$, with $n\in {\bbb Z}^+$ and $n\geq
\ell$.
\end{enumerate}

Therefore, as there is a correspondence between the results of the 
conventional and our factorizations, one might conclude the total
equivalence of both treatments. However, we make a remark worth to take
into account: the conventional factorizations make use of two Hamiltonian
hierarchies,  
${ H}^{\ell}_x$ and ${ H}^{\ell}_z$, whose terms differ in a
constant $4\ell$, while the new factorizations use only one ${ H}^{\ell}$. If
we want that both factorizations $(a)$ and $(b)$ be valid inside the same
hierarchy it is necessary to adopt the properties of our approach in the
following sense: either  the operators
$X_{\ell}^\pm$ or
$Z_{\ell}^\pm$ (or both pairs) must change not only the quantum number 
$\ell$ but also $n$. In this way we have shown, by means of this simple
example, that the factorizations presented here prove to be quite
useful  providing directly a more natural viewpoint.

\subsection{Morse Potential}

In this case we have eigenvalue Schr\"odinger equations for the whole real 
line $x\in{\bbb R}$ with the potentials
\be
\label{O23}
V^{\ell}(x) = \left( \frac{\alpha}{2} \right)^2 
\left( e^{2\alpha x} -
2(\ell+1) \, e^{\alpha x} \right), \qquad \alpha>0, \ell\geq 0.
\ee
Often in the literature \cite{Landau} the Morse potentials are written  
$V(y) = A  \left( e^{-2\alpha y} - 2\, e^{-\alpha y} \right)$. This form can be
reached from (\ref{O23}) by a simple change of the variable
$x=-y+k$, with $e^{\alpha k}= \ell+1$.

The energy eigenvalues can be expressed as
\be
\label{meigenmorse}
E^{\ell}_{n} = -\frac{\alpha^2}4\, n^2,\qquad 
n = \ell -2\nu  > 0;\quad  \nu=0,1,2\dots
\ee
In order to have bounded states it is necessary the restriction
$\ell>0$; the critical value $\ell=0$ has in this respect an special limiting
character, and it is convenient to take it into account as we shall see later.
According to (\ref{meigenmorse}), the eigenfunctions $\psi_{n}^{\ell}$
are characterized by labels satisfying $n \leq \ell $; this means that the 
ground states will be defined through (\ref{xmasB}).

There are two new factorizations
\be
\label{otramas}
\frac{-e^{-\alpha x}}{\alpha^2} \left[ H^{ \ell} -
E^{\ell}_{n} \right] = { B}^i_{ n, \ell}(x){ A}^i_{n,\ell}(x)
 - \phi^i(n,\ell), \qquad i=1,2,
\ee 
with $\phi^i{(n,\ell)}$ given by
\be
\phi^1(n,\ell) = -\frac12(\ell +n +2),\qquad
\phi^2(n,\ell) = -\frac12(\ell-n+2),
\ee
and the action on the parameters $(n,\ell)$ for each factorization 
by the functions 
\be
(n,\ell)= F^1(n+1,\ell+1),\qquad
(n,\ell)= F^2(n-1,\ell+1).
\ee
The explicit form of the intertwining operators (\ref{otramas}) is 
\bea
&&\left\{
\begin{array}{l}
{ B}^{1}_{n,\ell}(x)=
\frac{e^{-\alpha x/2}}{\alpha}\frac{d\,}{dx} {+}\frac12 e^{\alpha x/2} {+}
\frac{n+1}2e^{-\alpha x/2} 
\\   [1.5ex] 
{ A}^{1}_{n,\ell}(x)= 
\frac{e^{-\alpha x/2}}{\alpha}\frac{d\,}{dx} {-}\frac12e^{\alpha x/2} {-}
\frac{n}2e^{-\alpha x/2} 
\end{array}
\right. 
\label{a}\\
&& \ \ \nonumber
\\
&&\left\{
\begin{array}{l}
{ B}^{2}_{n,\ell}(x)=
\frac{e^{-\alpha x/2}}{\alpha}\frac{d\,}{dx} {+}\frac12e^{\alpha x/2} {-}
\frac{n-1}2e^{-\alpha x/2} 
\\  [1.5ex]
{ A}^{2}_{n,\ell}(x)= 
\frac{e^{-\alpha x/2}}{\alpha}\frac{d\,}{dx} {-}\frac12e^{\alpha x/2} {+}
\frac{n}2e^{-\alpha x/2} 
\end{array}
\right.
\label{b}
\eea
As in the oscillator case we have two pairs of operators that change
simultaneously two types of labels: one, $\ell$, is related to the intensity of
the potential, although here it can not be interpreted as due to a centrifugal
term. The second one, $n$, is directly related to the energy through formula
(\ref{meigenmorse}). The (nonvanishing) commutators of the free-index 
operators are 
\be
\label{2bosonbix}
\begin{array}{lll}
[L,B^i] =  - B^i, \quad  & [N,B^i] =(-1)^iB^i,\quad
&[A^i,B^i] = 1
\\ [1ex]
[L,A^i] =  A^i, \quad & [N,A^i] = -(-1)^iA^i, \quad  & i=1,2. 
\end{array}
\ee

Observe that in this case the function 
$h_{n,\ell}(x) =  {-e^{-\alpha x}}/{\alpha^2}$ is not a constant, so the
hermiticity relations among the operators $\{A^i,B^i;i=1,2\}$ are spoiled. Let 
us take $\ell\in {\bbb Z}^+$, and formally allow for negative $n$-values in
(\ref{meigenmorse}), i.e., $\pm n = \ell -2\nu$; this is admissible because 
in the operators of (\ref{a})-(\ref{b}) we have a symmetry under the change
$n\to -n$. Then the Hilbert space $\cal H$ of bounded states enlarged with
the  (not square integrable) states $\psi_{n=0}^{\ell}$, $\ell=0,1,2\dots$, will
be invariant under the action of all the operators defined in this section. The
lowest weight state is played in this case by a not square-integrable
wavefunction,
$\psi_{n=0}^{\ell=0}$.

We can of course build other operators out of the previous ones, changing
exclusively one of the labels: the pair $\{A^1A^2, B^1B^2\}$ change $\ell$ (in
$+2$ or $-2$ units, respectively), while 
$\{A^1B^2, A^2B^1\}$ change $n$ (also in $+2$ or $-2$ units, respectively).
It is interesting to show explicitly the form taken by the former couple: 
\be
\left\{
\begin{array}{l}
{ (B^1B^2)}_{n,\ell}=
\frac1{\alpha}\frac{d\,}{dx}
{+}\frac12 \left( e^{\alpha x} {-} (\ell+2) \right)
\\  [1ex]
{ (A^1A^2)}_{n,\ell}= 
-\frac1{\alpha}\frac{d\,}{dx}
{+}\frac12 \left( e^{\alpha x} {-} (\ell+2) \right) \ ,
\end{array}
\right.
\ee
where $(A^1A^2)_{n,\ell}= A_{n-1,\ell+1}^1A^2_{n,\ell}$ and 
$(B^1B^2)_{n,\ell}=B^1_{n,\ell}B^2_{n+1,\ell+1}$, according to the rules of the
action of free index operators (\ref{2.2}), (\ref{free}).  They can be
identified with the usual factorization operators for the Morse Hamiltonians
$H^{\ell}$ described in the first section in the following way

\begin{enumerate}
\item
Factorization:  $X_{\ell'}^+ X_{\ell'}^- - q(\ell') = { H}^{2\ell'},\ \ell' \in {\bbb
Z}^+.$
\item
Operators: $X_{\ell'}^+ = -\alpha\,  (B^1B^2)_{n,2\ell'}$\ , 
$X_{\ell'}^- = -\alpha\,  (A^1A^2)_{n,2\ell'}$\ , 
$q(\ell')= \alpha^2 (\ell'+1)^2$\ .
\item
Ground states: $X_{\ell'-1}^+ \psi^{\ell'}_{\ell'}=0$, $\ell' > 0$.
\item
Energy eigenvalues: $ E^{2\ell'}_{2n'} = - \alpha^2
({n'})^2$, with $n= 2n'$, $ n' \in {\bbb Z}^+$, and $0\leq n'\leq \ell'$.
\end{enumerate}
This time the notation, as it was mentioned above, is in agreement with
(\ref{xmasB}).

\subsection{Radial Coulomb Potential}

After the separation of the angular variables, the stationary
radial Schr\"odinger equation for the Coulomb potential in two dimensions
takes the form
\be
\label{0.1}
H^{\ell} \psi_n^{\ell}(r) =\left\{ -\frac{d^2\,}{dr^2} +
\frac{(2\ell+1) (2\ell  -1)}{4 r^2} -\frac{2}{r}\right\} \psi_n^{\ell}(r) =
E^{\ell}_n \psi_n^{\ell}(r),
\ee
where the values of the orbital angular momentum are positive integers
$\ell=0,1,2\dots$

The computation of the discrete spectrum associated to the bounded states 
of  $H^{\ell}$ can be easily obtained by means of the conventional
factorizations (\ref{1.1}) with
\be
\label{3.1}
X^\pm_{\ell} = \mp\frac{d\,}{dr} -\frac{2\ell +1}{2r} +\frac2{2\ell+1},
\qquad
q(\ell) = \frac{-1}{\left(\ell +1/2\right)^2} \  .
\ee
Therefore, according to the results quoted in  Section I, we have
\be \label{en}
E_n^{\ell}= - \frac1{(n+1/2)^2},\quad n=\ell+\nu ,\quad \nu=0,1,\dots
\ee

When our method is applied to
the hydrogen Hamiltonians $H^{\ell}$ of equation (\ref{0.1}) with the
eigenvalues $E_n^{\ell}$ (\ref{en}), we obtain two
independent solutions that read as follows
\bea
&&{ B}^{1}_{n,\ell}{ A}^{1}_{n,\ell} +\ell +n +1=
-\frac{(2n+1)r}4 \, \left[ H^{\ell} - E_{n}^{\ell} \right] ,\\
&&{ B}^{2}_{n,\ell}{ A}^{2}_{n,\ell}  -\ell +n +1 =
-\frac{(2n+1)r}4 \, \left[ H^{\ell} - E_{n}^{\ell} \right].
\eea
The explicit form of the operators
$\{A^i,B^i\}_{i=1,2}$,
is displayed below:
\bea
&&\left\{
\begin{array}{l}
{ B}^{1}_{n,\ell}(r)=
(2n+1)^{1/2}
\left( \frac{r^{1/2}}{2}\frac{d\,}{dr} {+}\frac{r^{1/2}}{2n+1} {+}
\frac{\ell}{2r^{1/2}} \right)
c_n^{-1/2}\, D({c_n})
\\ [2ex]    
{ A}^{1}_{n,\ell}(r)= 
D({c_n}^{-1}) 
\, c_n^{1/2}
(2n+1)^{1/2}
\left( \frac{r^{1/2}}{2}\frac{d\,}{dr} {-}\frac{r^{1/2}}{2n+1} {-}
\frac{2\ell+1}{4r^{1/2}} \right)
\end{array}
\right. \label{Ca}\\
&&\nonumber\\[1ex]
&&\left\{
\begin{array}{l}
{ B}^{2}_{n,\ell}(r)=
(2n+1)^{1/2}
\left( \frac{r^{1/2}}{2}\frac{d\,}{dr} {+}\frac{r^{1/2}}{2n+1} {-}
\frac{\ell}{2r^{1/2}} \right)
c_n^{-1/2}\, 
D({c_n})
\\  [2ex]
{ A}^{2}_{n,\ell}(r)= 
D({c_n}^{-1}) \,
c_n^{1/2}
(2n+1)^{1/2}
\left( \frac{r^{1/2}}{2}\frac{d\,}{dr} {-}\frac{r^{1/2}}{2n+1} {+}
\frac{2\ell+1}{4r^{1/2}} \right)
\end{array}
\right.
\label{Cb}
\eea
The symbol $D(\mu)$ in (\ref{Ca})--(\ref{Cb}) is for the dilation
operator (\ref{dil}), and $c_n=\frac{2n+2}{2n+1}$. Thus, in this example we
are dealing with general first order differential operators as explained in
Section~I.  For the first couple
$\{A^1,B^1\}$ we have  $({\hat n},{\hat \ell}) =(n+1/2,\ell+1/2)$, while for
the second pair $\{A^2,B^2\}$, $({\hat n},{\hat \ell}) =(n+1/2,\ell-1/2)$. 
 
The nonvanishing commutators among the free-index operators are
\be
\begin{array}{lll}
[N,B^i] =\frac{-1}2 B^i,\quad  
&[L,B^i] = (-1)^i\, \frac12 B^i ,\quad   &[A^i,B^i]=I,
\\ [1ex]
[N,A^i] = \frac12 A^i,\quad 
&  [L,A^i] = -(-1)^i\, \frac12 A^i, \quad &\quad  i=1,2\ .
\end{array}
\ee
In other words, as in the previous examples, we have a set of two 
independent boson operator algebras. The problem with these operators is
that they change the quantum numbers $(n,\ell)$ in half-units, so that they
do not keep inside the sector of physical wavefunctions. To avoid this
problem we can build quadratic operators \cite{Liu} $\{{ A}^i{ A}^j,{ B}^i{
A}^j,{ B}^i{ B}^j\}_{i,j=1,2}$ satisfying this requirement; such second-order
operators close the Lie algebra $sp(4,{\bbb R})$ \cite{Alhassid}, which
includes the subalgebra $su(2)$ (whose generators connect eigenstates with
the same energy but different $\ell$'s). It is worth to write down these
quadratic operators:
\be
\left\{
\begin{array}{l}
(B^2 A^1)_{n,\ell}=
\frac{(2\ell +1)(2n+1)}{2}
\left( 
\frac12 \frac{d\,}{dr} {+}\frac{2\ell +1}{4r} {-} \frac{1}{2\ell +1}
\right) 
\\ [1ex] 
(B^1 A^2)_{n,\ell}=
\frac{(2\ell +1)(2n+1)}{2}
\left( 
-\frac12 \frac{d\,}{dr} {+}\frac{2\ell +1}{4r} {-} \frac{1}{2\ell +1}
\right) 
\end{array}
\right.
\ee
They constitute, up to global constants, the usual
factorization operators given in (\ref{3.1}): $X_{\ell}^+\propto
(A^1B^2)_{n,\ell}$, $X_{\ell}^-\propto (A^2B^1)_{n,\ell}$. Another subalgebra 
is $su(1,1)$ (relating states with the same
$\ell$ but different energies or $n$ values). Once included the negative 
$\ell$ values, as we did for the radial oscillator potential, the space  ${\cal
H}$ is the support for what it is called a `singleton representation' 
\cite{singleton} of $so(3,2)\approx sp(4,{\bbb R})$. There is one
lowest weight eigenvector 
$\psi_{n=0}^{\ell=0}\in \cal H$, from which all the representation space is
generated by applying raising operators.


\section{Conclusions and Remarks}

We have shown that a refinement of the factorization method allows us to
study the maximum of relations among the Hamiltonian hierarchies that the
conventional factorizations are not able to appreciate.
The operators involved obbey commutation rules which show the
connection existing among the three examples dealt with in this paper: they
have the same underlying Lie algebra associated with confluent
hypergeometric functions. In other occasions the conventional factorizations
have been used in this respect, but we have seen that such an approach is
partial and not complete at all.

Usually the Hamiltonian hierarchies are obtained from higher dimensional 
systems after separation of variables (or by any other way of reduction).
Such systems have symmetries that are responsible for their analytical
treatment. These symmetries are reflected in the many factorizations that
the hierarchies can give rise to by means of the thecnique we have
developed. We have limited our study to $N=2$ space dimensions for the
radial oscillator and  Coulomb potentials because they are the simplest cases
to deal with. For other dimensions there appear certain subtleties, in the
sense that the Hilbert space ${\cal H}$ of bounded states is no longer
invariant under the involved operators
\cite{Black}.

Finally, let us mention that we have limited ourselves to some examples  (all
of them inside the class of shape invariant potentials \cite{Balantekin}), but
it is clear that the whole treatment is aplicable to the remaining
Hamiltonians in the classification of Infeld and Hull \cite{Infeld}.

\subsection*{Acknowledgements}

This work has been partially supported by a DGES project
(PB94--1115)  from Ministerio de Educaci\'on y Cultura (Spain), 
and also by Junta de Castilla y Le\'on (CO2/197). ORO acknowledges
support by SNI and CONACyT (Mexico), and the kind hospitality at the
Departamento de F\'{\i}sica Te\'orica (Univ. de Valladolid).


\end{document}